\documentclass{iau}
\usepackage{graphicx}

\title[IAUS 295.~~Assembly Histories of Simulated ETGs] 
{Assembly Histories and Observational Properties of Simulated Early-type Galaxies}

\author[Peter H. Johansson]   
{Peter H. Johansson$^1$}

\affiliation{$^1$Department of Physics, University of Helsinki, \\ Gustaf H\"allstr\"omin katu 2a,
FI-00014 Helsinki, Finland \\ email: {\tt Peter.Johansson@helsinki.fi} \\[\affilskip]}

\pubyear{2013}
\volume{295}  
\pagerange{xx--xx}
\jname{The intriguing life of massive galaxies}
\editors{D. Thomas, A. Pasquali \& I. Ferreras, eds.}
\begin{document}

\maketitle

\begin{abstract}
We demonstrate that massive simulated galaxies assemble in two phases, with the 
initial growth dominated by compact in situ star formation, whereas the late growth is dominated by accretion of old stars formed in 
subunits outside the main galaxy. We also show that 1) gravitational feedback strongly suppresses late star formation in 
massive galaxies contributing to the observed galaxy colour bimodality that 2) the observed galaxy downsizing can be explained naturally in the two-phased model
and finally that 3) the details of the assembly histories of massive galaxies are directly connected to their observed kinematic properties.
\end{abstract}

\firstsection 
\section{The two phased formation of early-type galaxies}

High-resolution numerical zoom-in simulations of massive early-type galaxies have shown
that there are two distinct phases in their formation histories 
(\cite[Naab et al. 2007]{2007ApJ...658..710N}; \cite[Oser et al.  2010]{2010ApJ...725.2312O}; 
\cite[Lackner et al.  2012]{2012MNRAS.425..641L}).
At high redshifts of $z\sim 3-6$ the galaxies assemble rapidly through compact $(r<r_{\rm eff})$ in situ star formation. The later growth
of the galaxies proceeds predominantly through the accretion of stars formed in subunits 
outside the main galaxy. The majority of the accreted stars are added to the outskirts of the galaxies 
at larger radii $(r>r_{\rm eff})$ providing a satisfactory explanation for the observed size growth of massive
early-type galaxies (ETGs) since $z\sim 3$ until the present-day 
(\cite[Naab et al.  2009]{2009ApJ...699L.178N}; \cite[Bezanson et al.  2009]{2009ApJ...697.1290B}; 
\cite[Oser et al.  2012]{2012ApJ...744...63O}).

Here we show that in addition to the size growth of ETGs, the two phased formation picture can also
shed light on other observational results of ETGs, such as the observed galaxy bimodality, the downsizing of massive galaxies
and the observed dichotomy of the kinematic properties of ETGs. The simulation sample
we use to address these questions contains 9 galaxies simulated at high-resolution (\cite[Johansson et al. 2012]{2012ApJ...754..115J}) 
using the Gadget-2 code (\cite[Springel 2005]{2005MNRAS.364.1105S}).
The simulations include cooling for a primordial composition, star formation and feedback from type II supernovae, but exclude supernova driven winds and AGN feedback. 
We run three models (A2,C2,E2) at very high spatial ($\epsilon_{\star}=0.125 \ \rm kpc$) and mass resolution 
($m_{\star}\sim 10^{5} M_{\odot}$) with the remaining six simulations (U,Y,T,Q,M,L) simulated at a somewhat lower resolution of 
($\epsilon_{\star}=0.25 \ \rm kpc$, $m_{\star}\sim 10^{6} M_{\odot}$).
We organize the galaxies in our simulation sample into three groups of three galaxies each depending on 
whether their late assembly history $(z\lesssim 2)$ is dominated primarily by dissipationless minor merging
(mostly accreted stars: galaxies C2,U,Y), a mixed dissipationless/dissipational (mostly accreted
and some in situ stars: galaxies A2,Q,T) or a primarily dissipational formation history 
(significant amount of in situ stars: galaxies E2,M,L). 
We find that this classification of the assembly histories of ETGs is useful as it broadly separates
more massive ETGs forming dissipationlessly from less massive ETGs with a 
more dissipational formation history.

\begin{figure}[h]
\begin{center}
 \includegraphics[width=14.0cm]{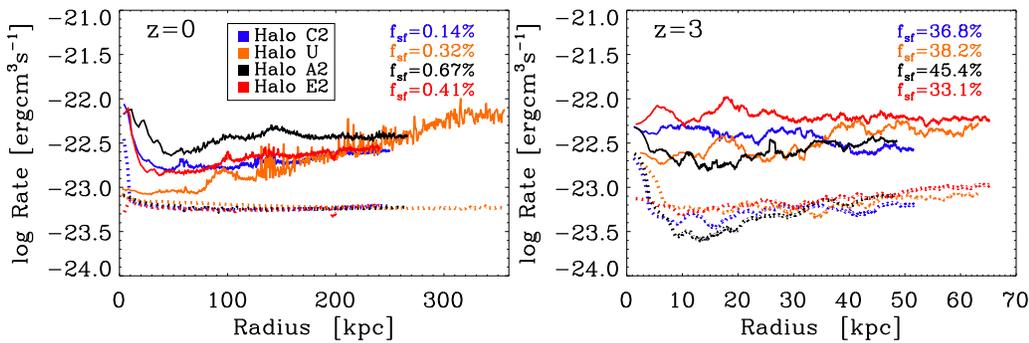} 
\caption{The net heating (solid line) and net cooling rates (dashed lines) for
  virial $(r<r_{\rm vir})$ non-starforming diffuse gas, for which the density is below the star formation
threshold of $n_{\rm th}<0.205 \ \rm cm^{-3}$ shown at redshifts of $z=0$ (left) and
$z=3$ (right).  The fraction $f_{\rm{sf}}$ of dense starforming gas $(n>n_{\rm th})$ is also given.
Typically, the heating rate dominates over the cooling rate at all redshifts for the low-density non-starforming gas.}
   \label{fig1}
\end{center}
\end{figure}

\section{The bimodality of the local galaxy population}

Recent large-field sky surveys have unequivocally demonstrated that local galaxies show a bimodal distribution.
Local galaxies with stellar masses above a critical mass of $M_{\rm crit}\simeq 3\times 10^{10} M_{\odot}$ are typically red
spheroidal galaxies with old stellar populations, whereas galaxies below this critical mass are typically blue, star-forming
galaxies with somewhat younger stellar populations. The observed bimodality is usually explained theoretically using models
in which the star formation is efficiently quenched in massive haloes above $M\sim 10^{12} M_{\odot}$  
(\cite[Dekel \& Birnboim 2006]{2006MNRAS.368....2D}; \cite[Gabor et al. 2011]{2011MNRAS.417.2676G}).

The late assembly of our simulated galaxies is dominated by dry minor merging building up the accreted stellar component.
The infalling satellite galaxies captured through dynamical friction will cause gaseous wakes from which
energy is transferred to the surrounding gas. Collectively together with the heating caused by supersonic
collisions and shocks caused by infalling cold gas this process is deemed gravitational heating. In Fig. \ref{fig1} we compare the shock-induced 
heating rates of the diffuse non-starforming gas with the corresponding cooling rates. 
The heating rates dominate over the cooling rates at all redshifts with the more
massive galaxies (U,A2) showing higher heating rates than the slightly lower mass galaxies (C2,E2), as expected by the scaling of
the gravitational feedback energy,  $(\Delta E)_{\rm grav}\propto v_{c}^{2}$, where $v_{c}$ is the circular velocity of the galaxy
(\cite[Johansson et al. 2009b]{2009ApJ...697L..38J}).
The inclusion of gravitational heating helps in maintaining a hot gaseous halo and thus inhibits star formation contributing to the 
observed galaxy bimodality. However, gravitational heating is predominantly important in the outer parts of galaxies and some form of additional feedback, 
most probably AGN feedback (e.g. \cite[Johansson et al. 2009a]{2009ApJ...690..802J}), is required to stop late central star formation 
in very massive galaxies (e.g. galaxy U at $z=0$).

\section{Downsizing of massive galaxies}

Several recent observations have shown that old, massive red metal-rich galaxies were already
in place at high redshifts of $z\sim 2-3$. This observational result can be seen as a manifestation
of cosmic downsizing, in which galaxies seem to form anti-hierarchically in the sense that the most massive
galaxies formed a significant proportion of their stars at high redshifts, compared to lower mass 
systems that exhibit a more continuous star formation history throughout the cosmic epoch 
(e.g. \cite[Glazebrook et al. 2004]{2004Natur.430..181G}).

\begin{figure}[t]
\begin{center}
 \includegraphics[width=14.0cm]{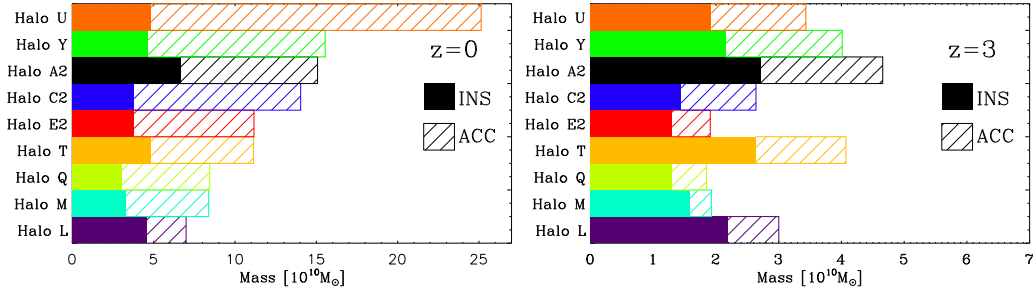} 
 \caption{The stellar mass assembly histories of our simulated galaxy sample, with
the solid colour bars showing the contribution of in situ formed stellar mass and
the dashed colour bars representing the contribution of accreted stellar mass shown
at redshifts of $z=0$ (left) and $z=3$ (right). At high redshifts ($z\gtrsim 3$) 
the galaxies assemble rapidly through in situ star formation, whereas the late  
($z\lesssim 3$) assembly history is dominated by accreted stars, with the more massive
galaxies ending up with a proportionally larger fraction of accreted stars.}
   \label{fig2}
\end{center}
\end{figure}

In Fig. \ref{fig2} we show the assembly histories of all our simulated galaxies
depicting separately the masses in the in-situ formed and accreted stellar components. At $z=3$ the stellar
components in all galaxies have been assembled rapidly through mainly in situ star formation, fueled 
by cold gas flows and hierarchical mergers of multiple star-bursting subunits. At lower redshifts 
 $(z\lesssim 3)$ the subsequent growth of the stellar component proceeds predominantly through the
accretion of existing stellar clumps. The galaxies in Fig. \ref{fig2} are ordered
in decreasing final stellar mass from top to bottom and we can immediately see that the fraction of accreted
stars at $z=0$ increases as a function of galaxy mass. The accreted stellar component forms on average in low mass galaxies very early in
the Universe. However, the accreted stellar component is added 
to the massive galaxies much later and at lower redshifts than the in situ formed stars. In addition, the metallicity of
the central in situ formed stars is on average higher than the metallicity of the accreted stars that are formed in lower mass galaxies
and added later to the outskirts of galaxies, resulting in a negative metallicity gradient, in agreement with the observations. Thus, the counter-intuitive concept of
downsizing can be explained in the two phased formation mechanism. Massive galaxies form their central stellar mass in situ and then
accrete substantial amounts of stars that were formed even earlier in smaller subsystems. Hence by $z\sim 2-3$ the most massive galaxies 
have the oldest stellar populations compared to lower mass galaxies that are still forming in situ stars.

\section{Kinematic properties of massive galaxies}

\begin{figure}[t]
\begin{center}
 \includegraphics[width=14.5cm]{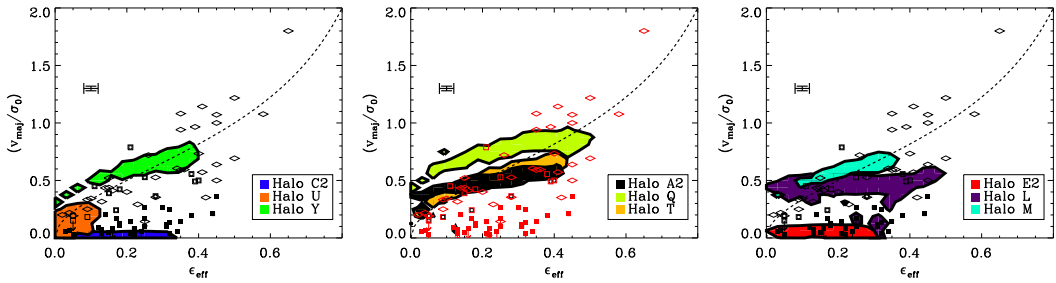} 
 \caption{The effective ellipticity $(\epsilon_{\rm eff})$ is plotted against the ratio of the 
major axis rotation and central velocity dispersion $(v_{\rm maj}/\sigma_{0})$. The contours show the 95\% probability
location of the galaxies in the $(\epsilon_{\rm eff}-v_{\rm maj}/\sigma_{0})$-plane derived from 500 random viewing
angles of our simulation data. The overplotted symbols are observational data from (\cite[Bender et al.  1994]{1994MNRAS.269..785B})
demonstrating that our simulated galaxies seem to be largely consistent 
with the observations. The dashed line shows the theoretical value for an oblate isotropic rotator.}
\label{fig3}
\end{center}
\end{figure}

Observations in the late 1980s using  slit spectroscopy found that the ETG population can be broadly separated into a 
population of massive slowly-rotating systems $(v/\sigma<0.1)$ 
with boxy isophotes and a population of fast-rotating $(v/\sigma\sim 1)$ ETGs with more disky isophotes found typically at somewhat lower masses.
Recent results from the volume limited ATLAS$^{\rm 3D}$ survey utilizing a modern integral-field-unit (IFU) confirmed this dichotomy
showing that about $\sim 15\%$ of local ETGs rotate slowly with no indications of an embedded disk component, whereas the 
majority ($\sim 85\%$) of the local ETGs show significant disk-like rotation (\cite[Cappellari et al. 2011]{2011MNRAS.413..813C}).

We derive the kinematic properties of our simulated galaxies using 500 random projections in order to assess the mean properties of
the simulated galaxies averaged over all sightlines. In Fig. \ref{fig3} we plot the 95\% probability of finding a simulated galaxy  
in the $\epsilon_{\rm eff}-(v_{\rm maj}/\sigma_{0})$ plane, where $\epsilon_{\rm eff}$ measures the effective ellipticity of the galaxies and
the ratio of the major axis rotation and velocity dispersion $(v_{\rm maj}/\sigma_{0})$ is a measure of the rotational support of the galaxies. 
Galaxies U,C,E2 show projections with very low rotational support, whereas the other galaxies (Y,A2,Q,T,L,M) 
show significantly more rotational support in very good agreement with the theoretical prediction for an oblate isotropic rotator shown 
by the dashed line in Fig. \ref{fig3}. The most
massive galaxy U shows mostly round projections with very low $\epsilon_{\rm eff}$, whereas all the other galaxies show projections extending
in $\epsilon_{\rm eff}$ from zero up to 0.4.
We see a correlation between the rotational support and the in situ/accreted fraction, both galaxies U,C that have
a large accreted fraction are slowly rotating and galaxy Y would most probably also been slowly rotating if it had not experienced 
a late $(z<0.5)$ major merger. Thus, within our rather narrow mass range the majority of the simulated galaxies are consistent 
with being rotationally supported disky ellipticals, whereas the most massive galaxy in our sample is consistent with being a roundish slow-rotator.

\end{document}